# Scientific Computing in the Cavendish Laboratory and the Pioneering Women Computors

C. S. Leedham[a]* and V. L. Allan[b]†

*aInstitute of Astronomy, University of Cambridge, Cambridge, UK; bCavendish Laboratory, University of Cambridge, Cambridge, UK*

*Contact: cscl3@cam.ac.uk

†Contact: vla22@cam.ac.uk

Caitriona Leedham is a student at the Institute of Astronomy, University of Cambridge. Research interests include the history of radio astronomy and women in STEM. Caitriona's ORCID ID is https://orcid.org/0000-0001-6879-7667.

Verity Allan is an Engineer Co-ordinator at the Cavendish Laboratory, University of Cambridge. Research interests include HPC for radio astronomy, the history of radio astronomy, computing, and women in STEM. Verity's ORCID ID is https://orcid.org/0000-0003-1625-5991 and you can get in contact on Twitter @verityallan.

# Scientific Computing in the Cavendish Laboratory and the Pioneering Women Computors

The use of computers and the role of women in radio astronomy and X-ray crystallography research at the Cavendish Laboratory between 1949 and 1975 have been investigated. We recorded examples of when computers were used, what they were used for and who used them from hundreds of papers published during these years. The use of the EDSAC, EDSAC 2 and TITAN computers was found to increase considerably over this time-scale and they were used for a diverse range of applications. The majority of references to computer operators and programmers referred to women, 57% for astronomy and 62% for crystallography, in contrast to a very small proportion, 4% and 13% respectively, of female authors of papers.

## Introduction

The three decades following the Second World War were incredibly exciting for scientists. Discoveries from war-time advances moved into the civilian sphere, kick-starting advances in computing and other disciplines. This was the case in the University of Cambridge, with advances in computing made by the Mathematical Laboratory, and scientific discoveries leading to Nobel Prizes for members of the Cavendish Physical Laboratory.[1][2] Many of the advances in physics depended on the use

---

[1] Haroon Ahmed, *Cambridge Computing: The First 75 Years* (Third Millenium, 2013), *passim.*

[1)]   [2] Malcolm Longair, *Maxwell's Enduring Legacy: A Scientific History of the Cavendish* (Cambridge: Cambridge University Press, 2016), *passim.*

of the computers of the Mathematical Laboratory.[3]

One of us (CS) received a grant to study how computers were used by Cavendish researchers. We also looked at who was acknowledged as participating in these discoveries, and we looked at the gender of those acknowledged as authors, computer programmers and operators. We specifically looked at the gender of those acknowledged, because of the perception that physics and computing were and are male-dominated fields. Hicks notes that a "great man" history of computing leaves out many contributors; the same is true of physics.[4] The role of women in these fields is steadily being uncovered,[567] and we were keen to illuminate the role of women in the work of the Cavendish Laboratory during the years 1949-1975, when the University first had access to computers, and when the Cavendish was doing Nobel-winning science, often using those computers.

---

[2] [3] P. Chris Broekema, Verity L. Allan and Henri E. Bal, 'On Optimising Cost and Value in eScience: Case Studies in Radio Astronomy', IEEE (2018) < https://doi.org/10.1016/j.ascom.2019.100337> (section 7.1)

[3] [4] Mar Hicks, *Programmed Inequality: How Britain Discarded Women Technologists and Lost Its Edge in Computing* (Cambridge, MA: MIT Press, 2017), pp 4-5, p. 19.

[4] [5] Margot Lee Shetterly, *Hidden Figures: The American Dream and the Untold Story of the Black Women Mathematicians Who Helped Win the Space Race* (New York: William Morrow, 2016), *passim.*

[5] [6] Bodleian Libraries, *Oxford Women in Computing: An OralHistory*, podcast, 2020, <https://podcasts.ox.ac.uk/series/oxford-women-computing-oral-history> [accessed 20 July 2021]

[6] [7] Claire Evans, *Broad Band: The Untold Story of the Women Who Made the Internet* (New York: Penguin Putnam Inc, 2018), *passim.*

This is not an intersectional analysis.[8] The data that are recoverable at this point do not include social class, sexuality, race or disability; these were not routinely collected by the University during that time period. Nevertheless, we expose the contribution of a minoritized group in science and computing, and also show how the use of computers became normalized in physics.

Our analysis was focused on radio astronomy and X-ray crystallography. Both are known from personal communications to have used methods requiring the Fourier Transform, which involves decomposing a signal, such as from an array of telescopes or light diffracted off a crystal, into individual frequencies. This requires large sums of complex exponentials, so was likely to have been done using computers in order to speed up analysis and allow handling of greater volumes of data. To do these calculations by hand, astronomers and astrophysicists used Beevor-Lipson strips, but this was still a complex and error-prone computation.[9] This made these domains a good place to start investigating the use of electronic computers. The use of computers allowed crystallographers to image larger and more complex molecules, and was required for the interferometers designed by the Radio astronomy Group.[10][11]

Research in the Cavendish laboratory, the physics department of the University of Cambridge, during the 1950s to 1970s in both radio astronomy and X-ray crystallography was of considerable importance.[12] Several Nobel prizes were earned, including one in 1974 by Martin Ryle and Anthony Hewish for their pioneering research in radio astronomy, particularly around telescope design.[13] It was also during this time that the EDSAC 1, EDSAC 2, TITAN and IBM computers became operational at the University Mathematical Laboratory in Cambridge (1949, 1958, 1964 and 1971 respectively).[14] The EDSAC 1 and 2 computers were built by Maurice Wilkes and his team and were unique to Cambridge. The EDSAC was the University's first machine to electronically store programs, the second of those in the world. The EDSAC 2 was replaced by TITAN, a prototype for Ferranti's Atlas 2 computer.[15] TITAN was a time-sharing system, which allowed it to be used from multiple terminals in different locations simultaneously. The University then bought an IBM 370/165, its first commercial computer.

---

[10] Longair, *Maxwell's Enduring Legacy*, p. 379.

[11] P. A. G. Scheuer, 'Aperture Synthesis at Cambridge', p.250.

²⁾ [12] Longair, *Maxwell's Enduring Legacy*, pp. 414-420.

[13] Longair, *Maxwell's Enduring* Legacy, p. 285.

[14] Longair, *Maxwell's Enduring* Legacy, pp. 45-56, 57-62, 66, 87.

⁹⁾ [15] Simon Lavington, 'The Atlas Story', *Atlas Symposium*, 6 (2012), p. 13.

The development of computers was a major step in the advancement of many scientific disciplines. Historical research of computers in Cambridge and their involvement in astronomy and crystallography has shown how important they were to each field.[16] We demonstrate this importance quantitatively by using papers published at the time by members of the Cavendish from the relevant research groups. Our other focus of investigating the impact women had at the Cavendish during this time of scientific importance was also done using publications from the time.

The next section will describe how computer use and other elements of the research were quantified. Observations we made and graphs of the information we collected are in the results section, followed by a discussion and finally a conclusion.

**Methods**

***Radio Astronomy Papers***

The information collected in this investigation came solely from papers published by members of the Cavendish during these years. These were read to find any mentions of computers or the role of women.

The sample of radio astronomy papers was made up of all papers in a premade bibliography from the Radio Astronomy Group from 1949 up to the end of 1975. This date was chosen to end the range of investigation as TITAN was decommissioned in 1974 and this captures some of the "tail" of papers published after its official decommissioning date (see Figures 1 and 5 in the results section). The papers were read through and information on computer use, authors and notable acknowledgements, telescope use and inferred use of computers was recorded.

---

[16] David J. Wheeler, 'Programmed Computing at the Universities of Cambridge and Illinois in the Early Fifties', in *A History of Scientific Computing*, ed. by Stephen G.Nash (New York, ACM Press, 1990), pp.269-279 (p. 276).

Computer use includes any mention of a computer or computer program being used. This ranged from naming the computer and describing the calculations it was performing to only an acknowledgment at the end to the University Mathematical Laboratory, which ran the computers. While the word *computer* originally referred to women, because electronic computers were often referred to by name, especially in the early days, it is easy to distinguish electronic computing from computing by humans. A fairly typical example follows:

> Through the kind cooperation of the Director of the Cambridge University Mathematical Laboratory the computing was handled by the EDSAC-I, a large electronic computer. A programme was devised which performed the entire computation in 15 hours.[17]

The notable acknowledgements collected information about both computer use and the role of women in the department. Any mention of who operated or programmed the computers was included as well as references to contributions by women for other reasons. Gender could usually be inferred by titles, but if not, the individual people were found in biographies or staff lists online. Since most of the titles are from these papers from the 1950s-1970s, this gives us a very binary view of gender. This is necessarily somewhat imprecise but allows capture of sufficiently accurate data for our purposes.

---

[17] J.H. Blythe, 'A New Type of Pencil Beam Aerial for Radio Astronomy', *Monthly Notices of the Royal Astronomical Society*, 117 (1957), 644-651 (p. 650).

Information about authors was found in a similar way. All papers considered were published under the name of at least one Cavendish scientist, but every author of those papers, regardless of institute, was recorded in the examination of gender distribution.

The sections on telescope use and inferred uses of computers were added after it became clear that as computers were becoming more common, authors were no longer making note of their use. Indications of computer use for radio astronomy include contour maps, data reduction and processes requiring a large amount of calculations, which have been done in previous papers on a computer, such as Fourier analysis and numerical integration. The data used in these papers were collected at the Mullard Radio Astronomy Observatory, which made many advancements in its telescopes over the time period being studied. The new telescopes, such as the One-Mile telescope (operational from 1964)[18] produced a much larger amount of information than previous interferometry methods. The output of the telescopes was in the form of punched tape, which was processed by a computer. This involvement of a computer was rarely stated but was instead inferred from the use of a telescope which we know from earlier papers required computer use for data preparation and processing.

---

9)    [18] Longair, *Maxwell's Enduring* Legacy, p. 414.

From this data, we tallied how many times the EDSAC, EDSAC 2, TITAN and IBM (specifically the one installed in Cambridge in 1974) computers were used as well as a count for unnamed computers and inferred uses. For a paper to be included in the inferred use category, it was required to include implications of computer use only. Papers including both implicit and explicit mentions of computer use were not counted in the implicit use category.

"Unnamed computers" included times when someone was acknowledged for computing work, but the computer wasn't mentioned; the University Mathematical Laboratory was thanked for its facilities, or a computer program was mentioned. It is possible that some of the earliest references include human computers, as the Mathematical Laboratory did perform calculations for people before electronic computers were invented. However, the numbers involved are very small, as it is much more usual to refer to a program. Indeed, human computation was often explicitly acknowledged, as in a 1953 paper by Ryle and Scheuer.[19] The inferred use column could have been included as an example of an unnamed computer, but we decided to keep these distinct, as this allowed us to infer information about how routine it was to use a computer in these scientific domains.

Other computers were mentioned when the Cavendish researchers collaborated with another institute and these were also tallied in a separate category, "other computers".

The total computer use is not a sum of all computer use categories. This would result in double counting as some papers used more than one type of computer. We counted each paper referring to computer use by one or more of the ways described above only once to obtain a tally of the total computer use for each year.

### X-Ray Crystallography Papers

The process for investigating papers in X-ray crystallography by the Cavendish was similar except that a list of papers had to be gathered first. This was done by starting with some well-known authors (primarily the Nobel prize winners from the Cavendish) and using some preliminary searches to find co-authors from their papers. After a list of people working at the Cavendish at the time was built up, searches were conducted by looking for papers by an author in one journal at a time. The journals used were based on the most frequently appearing titles during the preliminary searches. For authors who returned the most results, a search in Google Scholar was done for any papers published in other journals.

The list of papers was gathered in this way to minimize bias in areas of interest. For example, searching for papers from the Cavendish by keywords such as EDSAC would produce a set of papers which does not accurately represent how common the use of computers was overall. Also, we made the list first and then information was taken from the papers in one go to avoid changes in methodology over time.

Issues due to this method of paper collection which emerged in the data are considered in the discussion section.

Not all papers found could be used. Some searches returned papers not from the Cavendish as there were other places in Cambridge studying X-ray crystallography following restructuring of various university departments as new scientific fields opened up (see Figure 5 and the discussion section). Furthermore, some authors were also researching topics outside the field of crystallography. The definition of X-ray crystallography applied here was quite liberal and the papers include those concerning the development of the subject, but also the application of the technique for other scientific investigations. The searches were for publications between 1949 and 1975, to match the time period used for the Radio Astronomy papers.

When analysing the papers, there were some minor differences. There were fewer obvious indications for computer use when none was stated as looking for telescope use was naturally not relevant for crystallography, and there were no equivalent facilities that were stated or known from other sources to require computers as part of their routine operation. There were also many reports which mentioned how the use of computers was affecting the field of crystallography. These were noted down, but the final count of computer uses was limited to times where the computer was actually used. Finally, there were many mentions in the earlier papers of other machines used, such as Hollerith tabulators,[20] which were electromechanical machines used mainly by women known as "Hollerith girls"; see the Discussion section for further details.[21] These were included as other computers; thus this categorisation includes

---

10) [20] F. W. Kistermann, 'Hollerith punched card system development (1905-1913)', IEEE Annals of the History of Computing, 27 (2005), 56-66.

4) [21] Hicks, p. 47.

human computers. We have seen no records of the radio astronomy group using any Hollerith machines.

A qualitative comparison between radio astronomy and X-ray crystallography of how each field used computers is given in the results section.

**Results**

Around 700 papers in radio astronomy and 300 in X-ray crystallography from the Cavendish published between 1949 and 1975 are represented in the data below. All papers, search methods, tables and graphs were collected in a spreadsheet.[22]

Data was collected for both radio astronomy and X-ray crystallography in two main topics: information on computer use in the Cavendish and on the role of women. For comparison to the Figures, the years of operation of the computers should be noted: EDSAC 1949-1958, EDSAC 2 1958-1965, TITAN 1964-1973, IBM from 1971.

As well as a quantitative depiction of computer use, there are many interesting notes to be made from comparing how computers were used and talked about in radio astronomy and x-ray crystallography and how this changed over the years being studied. During the research, brief notes were made about how computers were used for the papers that mentioned them. These are given in the spreadsheet, but we will describe a few interesting points.

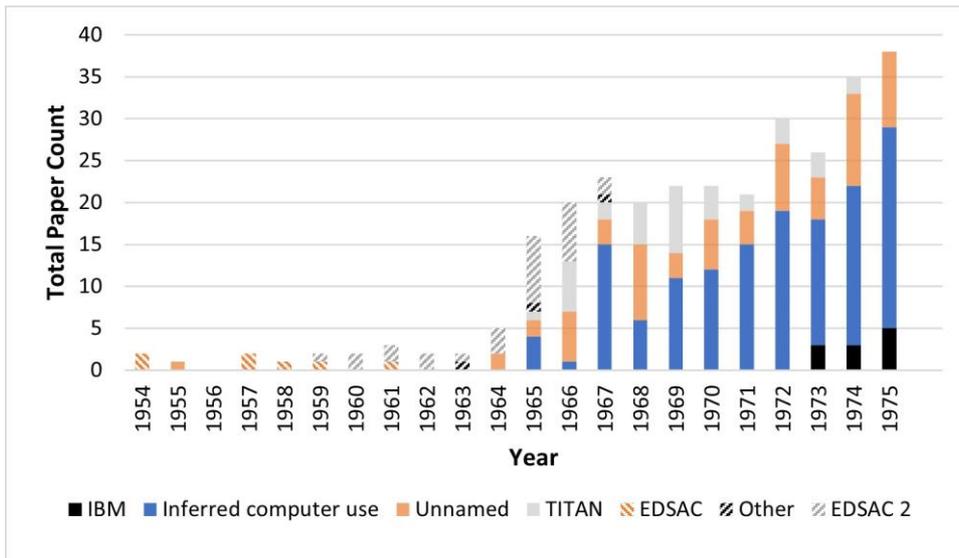

Figure 1: Computer use in radio astronomy papers. There is a general uptake in computers with a significant increase in 1965. The TITAN computer was built in 1964, but the dominant contribution in this year is due to the EDSAC 2. The increase is most likely due to the One-Mile telescope, built the same year as the TITAN. Different computers overlapped by about three years each time and a delay can be seen between the decommissioning of computers and when their use was published in a paper.

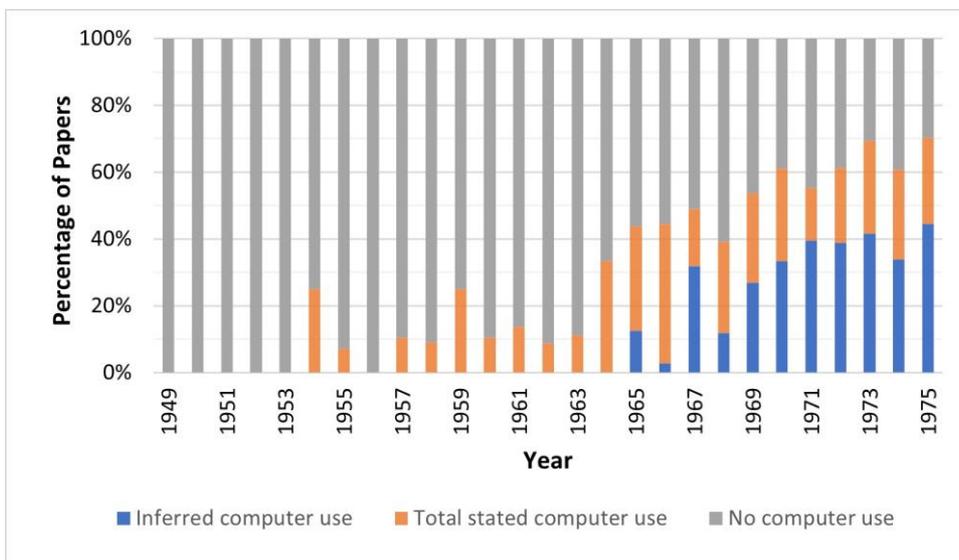

Figure 2: Total computer use in radio astronomy papers. The use of computers is shown as a proportion of the total number of papers. The proportion using computers, stated or inferred, increases until they are being used in the majority of papers. In the later years, more papers have implications of computer use rather than stating it. The inference of computer use starts in 1965, around the introduction of the One-Mile telescope.

During the early 1950s, both fields started off using computers in similar ways for performing large numbers of calculations that were not practical to do by hand. As expected, and part of the reasoning behind choosing radio astronomy and x-ray crystallography, Fourier computations and synthesis were the most frequent noted uses and remained so through to the 1970s. Other recurring uses include numerical integration and computations, least-squares refinement and error estimation.

Another shared use was for drawing contour plots. The change to using an automatic curve plotter seems to be earlier for crystallography, during the early 1950s, whereas an astronomy paper from 1962[23] mentions how EDSAC 2 sorted the output from convolution calculations so that contours could be easily drawn by hand.

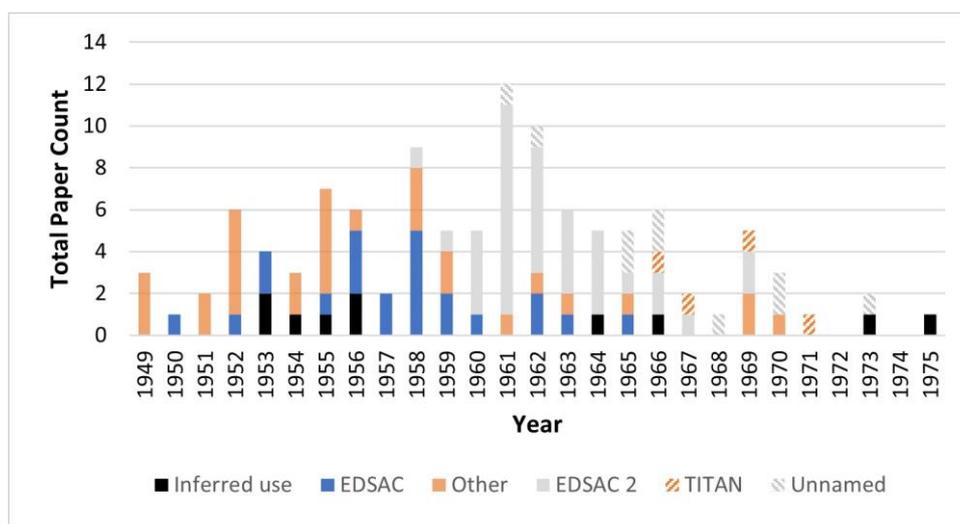

Figure 5: Computer use in X-ray crystallography papers. There is no clear trend in the amount of computer use. In 1962 when the LMB (Laboratory of Molecular Biology) opened, many crystallographers left the Cavendish. Considering up to 1962, there is a general positive trend, but it is less clear than for radio astronomy. Computers other than the Cambridge ones make a much more significant contribution. There is also a longer delay between the final acknowledgment of a specific computer and its decommissioning date. For example, the EDSAC was last mentioned in astronomy papers in 1961 compared to 1965 for crystallography. This reflects the longer time-scale over which X-ray crystallography research was conducted in this time period.

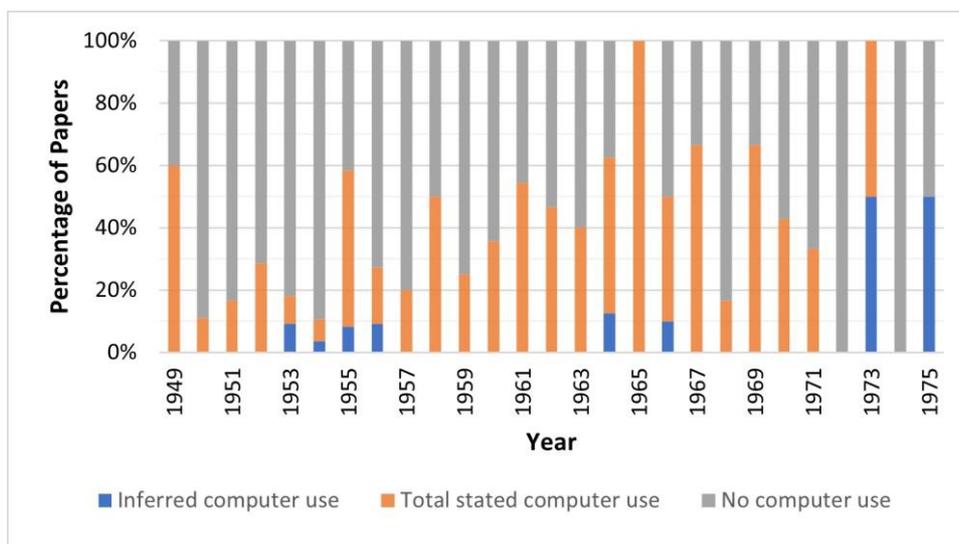

Figure 6: Total computer use in X-ray crystallography papers as a percentage of papers published in that year. There is no clear trend present here. The proportion of inferred computer use is much lower than in radio astronomy.

Crystallography was earlier to use more specialized programs, such as for computing atomic wave functions in 1953.[24] Reference to the actual program was also more common, such as the Busing-Levy program.[25] There were no named programs for radio astronomy, but certain standard computing methods could be seen by papers referring to previous ones detailing the same practices.

Overall, radio astronomy was quicker to omit any details of how they were using computers. This can be seen by comparing the inferred use of computers for the two fields in Figures 2 and 6.

Over the years, computers were starting the be used as more than just an efficient way to perform a lot of calculations. A 1965 crystallography paper simulated atomic orbits in a crystalline structure under standard force laws[26] and a 1975 astronomy paper described the results of a simulation of a disc of massless particles on a parabolic orbit around a potential well using a computer.[27]

We can also see from Figure 5 that within ten years, the use of Hollerith tabulators of various types was dominated by the use of electronic computers in crystallography. While it is not apparent from the diagram, the underlying data shows that the "other computers" then refer to electronic computers that were not run by the Mathematical Laboratory.[28].

---

There was a clear evolution of computer use from 1949 to 1975 in both fields. As more involved and specialized programs were created and the earlier techniques became standard practice, descriptions of how computers were being used became much less frequent.

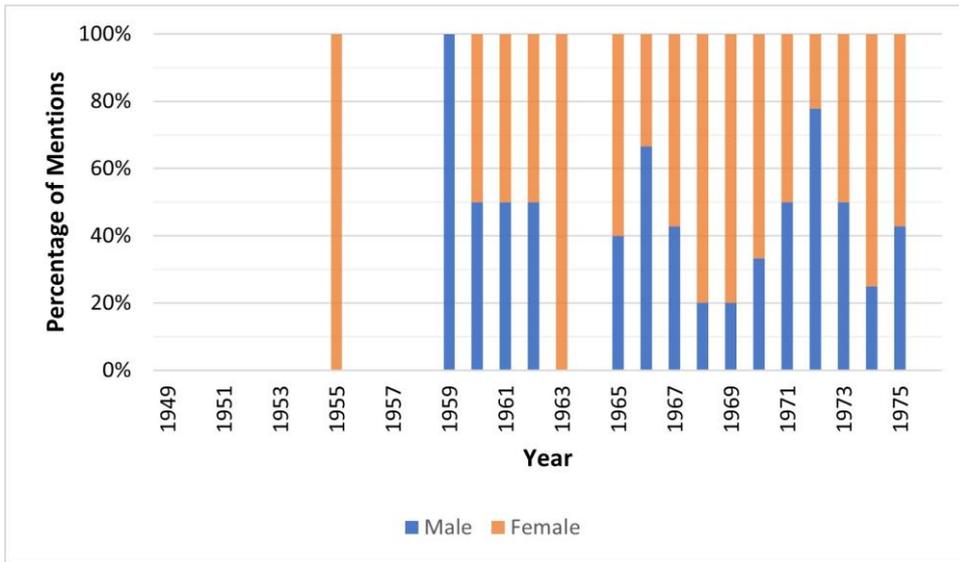

Figure 3: Gender distribution of computer programmers and operators in radio astronomy represented by how many times an individual was credited for computer use. Overall, women made up 57% of computer programmers and operators.

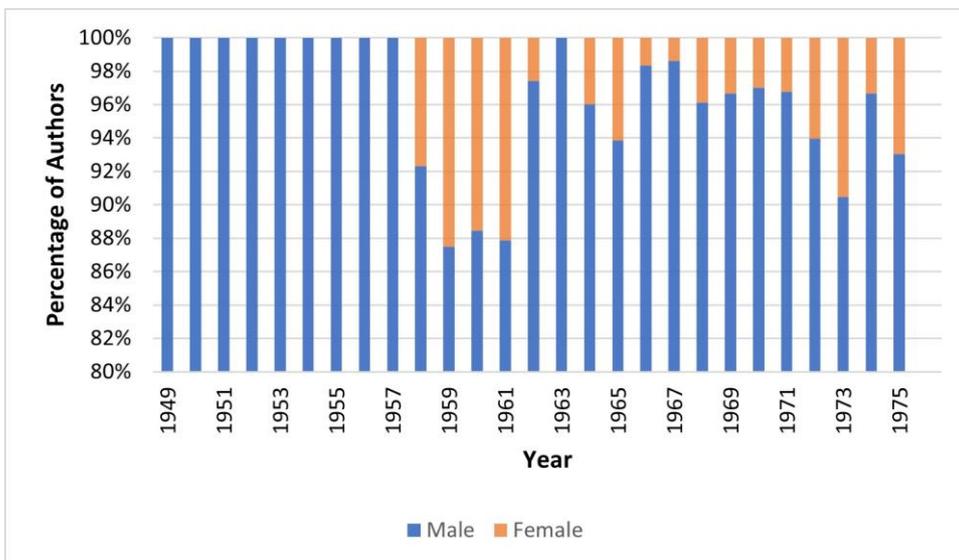

Figure 4: Gender distribution of authors in radio astronomy. Overall, women made up 4% of authors. There was less frequent representation for women as authors compared to computer operators. Neither this graph nor Figure 1 show an obvious trend, but all papers published by women were after 1958. The lack of obvious trends is in part because, although there was a large sample of papers, there were so few female authors that one on their own has a large impact on the data.

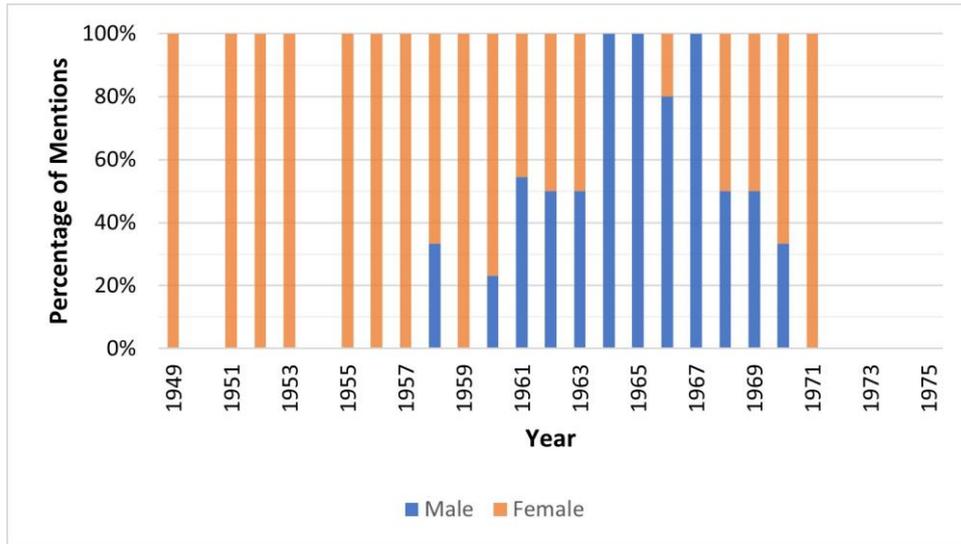

Figure 7: Gender distribution of computer programmers and operators in X-ray crystallography. Overall, women made up 62% of computer programmers and operators mentioned.

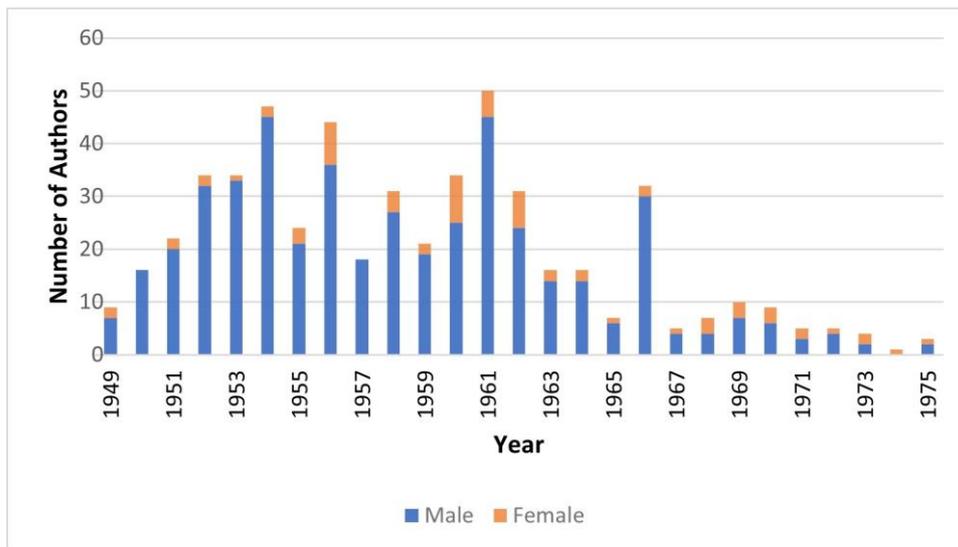

Figure 8: Gender distribution of authors in X-ray crystallography. Women made up 13% of the authors publishing the papers. Unlike for radio astronomy, these results are given as absolute values rather than percentages. The small number of papers found for the later years of the study leaves the gender balance unrepresentative and naïvely optimistic of the time.

**Discussion**

The aim of this research was to learn about how computers were being used and who was using them. The information in the results section demonstrates many interesting features of the evolution of computer use in radio astronomy and X-ray crystallography.

A simple comparison to make is the earliest reference found to a computer in each field. X-ray crystallography appears to introduce them much earlier. However most of those papers were not using the EDSAC, but machines such as the Hollerith tabulator, which were included in 'other'. Nevertheless, the first reference to the EDSAC was in 1950 for crystallography compared to 1954 for radio astronomy. These dates are very early, which was expected for these scientific disciplines based on methods they employed, such as the Fourier Transform, which are massively aided by using computers. First uses of computers in other branches of science should be investigated to provide a comparison.

The use of Hollerith tabulators provides a bridge between the era when computers were women, and the era when computers were electronic machines. Light observes that "A "computer" was a human being until approximately 1945. After that date, the term referred to a machine". Cortada provides more information about the range of Hollerith machines available. [29]It is not clear when this transition of definition occurred in the University of Cambridge, especially since Hollerith machines were used to confirm or augment crystallography results from the electronic computers for a number of years. This contrasts with the Mathematical Laboratory, which in the pre-war years, used Brunsviga, Monroe, and National calculating machines.[30] In this, the University was little different from most companies.[31]

One of the most significant differences between the fields was in inferring compute use. It is clear from Figures 2 and 6 that inferred use of computers is not as significant in crystallography as in radio astronomy. This is either because they were more explicit in crystallography or because of the way inferences were made. For radio astronomy, the main implication of computer use was telescopes. Crystallography didn't have an equivalent proxy for computer use. There were also far fewer graphs and diagrams that were computer drawn. Another contribution could be that there were

---

certain signs of computer use repeated across many papers that were missed here. We investigated X-ray crystallography second and kept the method applied similar to radio astronomy, so the same indicators were analysed, which may have missed several implications of the use of computers unique to crystallography.

Another limit to the conclusions that can be reliably drawn is the sample size of papers used for X-ray crystallography. As is clear in Figure 5, the departmental shift in 1962 does have a significant effect. This becomes most problematic in Figures 7 and 8 showing the gender distribution of computer operators and paper authors respectively as there is not enough data for the later years to interpret the graphs. Before 1962 the gender distribution is similar to that of radio astronomy in that the majority of people mentioned for computing work were women, but they only accounted for 13% of paper authors.

The small sample size also makes it difficult to draw strong conclusions from the near-parity between men and women acknowledged for computing. While we know from Abbate's work that women grew from a small minority of computer workers in the 1940s to 37% of those attaining computer science degrees in the 1970,[32] and there was a subsequent decline, and from Misa's work that women made up a minority of people in US professional societies and computing communities,[33] it is not clear whether a similar pattern holds in the UK. While women still make up a minority of those getting computer science degrees, in domains other than government work, where women seem to have been 30-40% of computer workers,[34] we have no data to show whether there was slow but steady growth (perhaps with periods of little to no growth), or whether there was a decline from a peak that was closer to parity. We are reluctant to assume that either the pattern that held in US industry or UK government work held in the world of university computing. While we have publicity photos, showing an all-women team of operators for EDSAC 2,[35] and recollections of majority-women teams,[36] it is dangerous to extrapolate this to the wider community of computer workers within the University.

Recollections from the period in question give us little indication about whether women were discriminated against, as happened to the women working in computing for the UK government.[37]  There is some indication that the Mathematical Laboratory, at least in the 1950s, like many other employers, would fire women on marriage or

---

pregnancy,[38] Radio Astronomy at least did not take that attitude.[39] When Roger Needham was running the Computer Laboratory (later to become the University's Department of Computer Science and Technology, one of the successors of the Mathematical Laboratory), he was proud to report that whenever women applied for an academic job there, they got it.[40] However, his tenure as Head of Laboratory was from 1980-1995, somewhat after the time considered here, and of course it does not indicate whether the women in service jobs were more likely to suffer discrimination.

Our data seems to fit more with Abbate and Misa's work that show that women were a small but significant minority of programmers.[41][42] What we can state is that there were women working in the field, and they were doing work that was significant enough to attract some recognition via inclusion in acknowledgements. As mentioned in the Improvements and Further Investigations section below, we were not able to investigate University records during the period of the grant, and rectifying this might allow us to shed more light on the gender distribution of University computer workers.

---

[38] Hicks, *Programmed Inequality*, p.10.

[39] Verity Allan, The Cavendish *Computors*: The Women Working in Scientific Computing for Radio Astronomy', (in preparation: accepted), pp.1-10, < https://doi.org/10.48550/arXiv.2205.07267 > [accessed 17 May 2022] (p. 8).

[40] Karen Spärck-Jones an oral history conducted in 2001 by Janet Abbate (IEEE History Center, Piscataway, 2001), <https://ethw.org/Oral-History: Karen_Spärck_Jones> [accessed 7 May 2022].

[41] Abbate, *Recoding Gender*, p. 68.

[42] Misa, 'Dynamics of Gender', pp. 76-77.

The results for the authors of crystallography papers are further complicated by the research method. Since it was done by searching for specific authors and there are a small number of papers for some years, the inclusion of papers from one particular author could have a sudden and significant effect. This could be resolved either by continuing the research to ensure all relevant papers are included or by a different approach to the search, which does not depend on author.

We were unable to draw conclusions about the programs used, and the software languages used, as these were not commonly discussed in the papers other than a general comment about the mathematical method, and there is limited archival material available for modern analysis.[43] This makes it difficult to trace the adoption of new programming languages in Cambridge, especially as EDSAC 2 used its own language. Wheeler suggested that this easy-to-use language actually hindered the adoption of other  programming languages.[44]

---

[43] Allan, *Cavendish Computors*, p. 8.

[44] David J. Wheeler, 'Programmed Computing at the Universities of Cambridge and Illinois in the Early Fifties', *A History of Scientific Computing*, ed. by Stephen G. Nash (New York, ACM Press, 1990), pp. 269-279, p. 278.

This makes it difficult to develop our history of scientific computing beyond the well-known machines and men that appear in existing histories.[454647] Preserving early programs does not seem to have been a priority, possibly because newer and better techniques were being devised all the time. Nevertheless, by focusing on the applications of computing, we hope that we have shown that computing formed an important, and later, essential part of science in the 1950s to 1970s.

### Improvements and Further Investigations

The set of radio astronomy papers used was assumed complete. For X-ray crystallography, it can only be assumed to be representative of the time and more papers could be found and added to the results. We were necessarily constrained by the time limit imposed by our grant funding.

Expanding the method of research used here to other departments in the Cavendish or to other scientific institutes could provide some context to the results and conclusions obtained here. The crystallography research in particular was affected by the opening of the Laboratory of Molecular Biology, so investigating crystallography in Cambridge outside of the Cavendish may be helpful. The range of data could also be expanded past 1975 to see how computing in Cambridge developed, though this may prove difficult, as we found explicit citations of computer use were declining by the 1970s.

---

[45] Mary Croarken, *Early Scientific Computing in Britain*, (Oxford, Clarendon Press, 1990), *passim*.

[46] *Scientific Computing: A Historical Perspective*, ed. by Bertil Gustafsson, (Cham, Springer, 2018), *passim*.

[47] *A History of Scientific Computing*, ed. by Stephen G. Nash (New York, ACM Press, 1990), *passim*.

To provide more detail into the role of women in computing, a distinction could be made between computer programmers and computer operators as the gender dynamics here might be different from the overall graphs given (Figures 3 and 7). With archival work being restricted during the period of our grant, we were unable to investigate University records, and the details provided in acknowledgements to articles are wholly insufficient to allow us to distinguish between the role of programmer and operator, if such a distinction even existed in practice. However, acknowledgements to the people using the computers were lacking compared to the amount of computer use, which may limit this. It may be more useful to find evidence of women in computing other than in the acknowledgments of papers, such as archived University records and staff lists. There were also staff working in other roles who were rarely acknowledged - the technicians who helped build instruments (mostly men) and the women who prepared diagrams and typed up scientific manuscripts were very rarely acknowledged, unless the work was considered exceptionally difficult. We have not yet seen a technician acknowledged in any of the papers we read for this study.

Thus the acknowledgements give us a very limited view of who contributed to a scientific paper, although it casts the net wider than those who were credited as authors.[48] We were limited in the resources we could access because of the Covid-19 pandemic, and once restrictions are lifted, there are more avenues that could be investigated. Nevertheless, looking at acknowledgements in papers exposes that there were more women involved in scientific discovery than can be uncovered by looking at lists of authors alone, and it is clear that these women and men were solving interesting and complex scientific and computational problems.

---

[48] Allan, *Cavendish Computors,* p. 4.

## Conclusions

An investigation using papers published by the Cavendish from 1949 to 1975 in radio astronomy and X-ray crystallography has shown an increase in computer use during this time. The EDSAC computers and TITAN became relied upon in many aspects of research over these years, such as in telescope use, map drawing and the vast quantities of calculations required to develop each field. Computers became more widely used and part of standard practice, appearing in most papers by 1975 and often no longer needing to be mentioned explicitly.

This study demonstrated that women played a major role in computing during these decades in contrast to their infrequent representation as authors of papers. While the overall numbers are small, they made up the majority of cited computer operators and programmers, which is made more significant when compared to the small proportion of women in computing today, and the small number of women authoring papers.[49] The results of this investigation show how computers became an essential element to radio astronomy and X-ray crystallography research at the Cavendish during the 1950s to 1970s and that the work of women in the use of these computers was indispensable.

---

8)  [49] S. Zweben and B. Bizot, '2019 CRA Taulbee Survey', *Computing Research News*, 32 (2020), 3-63, p. 6.

## Contributions

VA came up with the original project idea, supervised CL, and edited the paper. The astrophysics bibliography was compiled by members of the astrophysics group and maintained by Dr David Odell. CL compiled the data and the Crystallography bibliography and drafted the majority of the paper.

## Acknowledgements

We would like to thank Professor Malcolm Longair for his support of this project and Dr Elizabeth Waldram for interesting discussions. We thank Dr C.P Broekema and Dr D. Fenech for reading and commenting on this paper. CL would like to thank the Battcock Centre for Experimental Astrophysics for hosting her summer placement.

## Funding Information

CL received a grant from the UROP scheme and the Cavendish Laboratory.

## Disclosure Statement
We have no competing interests to declare.